\documentclass[conference]{IEEEtran}
\IEEEoverridecommandlockouts
\usepackage{cite}
\usepackage{amsmath,amssymb,amsfonts}
\usepackage{algorithmic}
\usepackage{graphicx}
\usepackage{textcomp}
\usepackage{xcolor}
\usepackage{soul}
\usepackage{subfigure}
\usepackage{multirow}
\usepackage{booktabs}
\usepackage{url}
\usepackage{caption}
\usepackage{balance}
\usepackage[shortlabels]{enumitem}
\newboolean{showcomments}
\setboolean{showcomments}{true}
\ifthenelse{\boolean{showcomments}}
{ \newcommand{\mynote}[2]{\textcolor{red}{
			\fbox{\bfseries\sffamily\scriptsize#1}
			{\small$\blacktriangleright$\textsf{\emph{#2}}$\blacktriangleleft$}}}}
{ \newcommand{\mynote}[2]{}}

\def \toolname {PSFinder}

\def\BibTeX{{\rm B\kern-.05em{\sc i\kern-.025em b}\kern-.08em
    T\kern-.1667em\lower.7ex\hbox{E}\kern-.125emX}}
    
\begin{document}

\title{Efficient Search of Live-Coding Screencasts from Online Videos\\}

\author{
    \IEEEauthorblockN{Chengran Yang, Ferdian Thung\textsuperscript{$\ast$}\thanks{$\ast$ Corresponding author.}, David Lo}
    \IEEEauthorblockA{\{cryang, ferdianthung, davidlo\}@smu.edu.sg}
}



\maketitle

\begin{abstract}
Programming videos on the Internet are valuable resources for learning programming skills. To find relevant videos, developers typically search online video platforms (e.g., YouTube) with keywords on topics they wish to learn. Developers often look for live-coding screencasts, in which the videos' authors perform live coding. 
Yet, not all programming videos are live-coding screencasts. In this work, we develop a tool named \toolname{} to identify live-coding screencasts. \toolname{} leverages a classifier to identify whether a video frame contains an IDE window. It uses a sampling strategy to pick a number of frames from an input video, runs the classifer on these frames, and then determines whether the video is a live-coding screencast based on frames classified as containing IDE window. In our preliminary experiment, \toolname{} can effectively identify live-coding screencasts as it achieves an F1-score of 0.97.


\end{abstract}


\section{Introduction}



\begin{figure*}[!htbp]
\captionsetup[figure]{font=normalsize}
\centerline{\includegraphics[width=\textwidth]{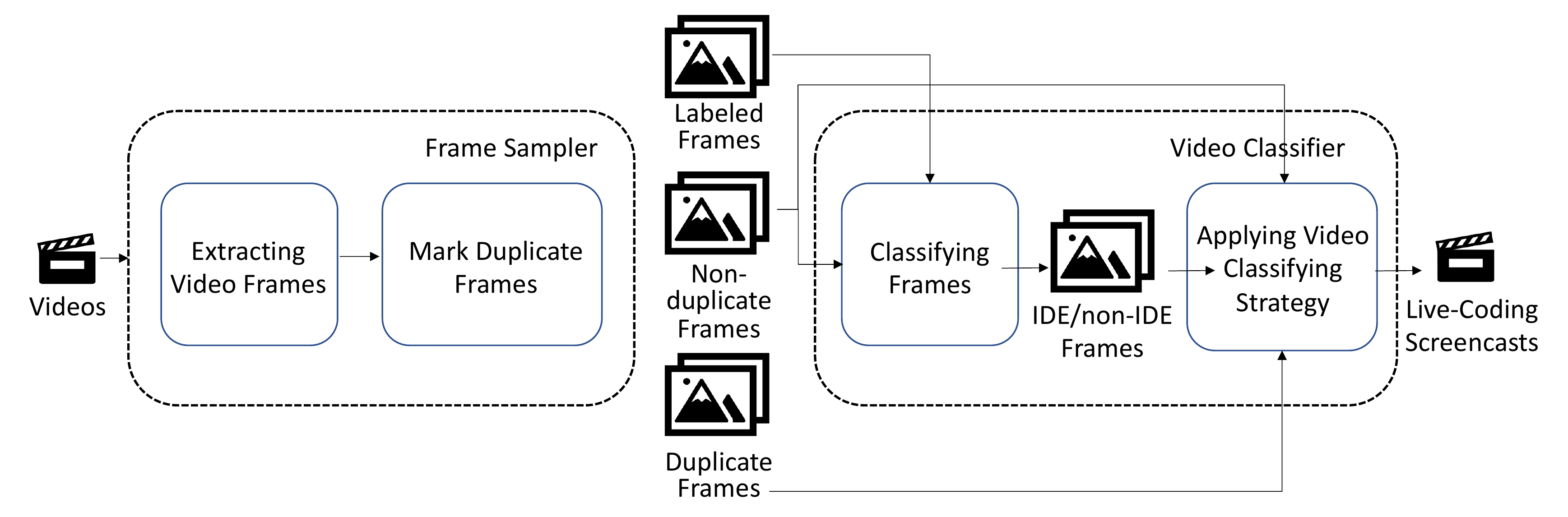}}
\caption{Overview of \toolname{}}\label{fig-approach}
\end{figure*}

Many live-coding screencasts are available on the Internet. For instance, the Massive Open Online Courses (MOOC) websites (e.g., Coursera\footnote{https://www.coursera.org/}), the nonprofit community that provides coding help (e.g., freeCodeCamp\footnote{https://www.freecodecamp.org/}), and millions of programming tutorials in YouTube\cite{bao2020psc2code} share their content in the form of live-coding screencasts.
In live-coding screencasts, we can see people programming in real-time, which guides them to quickly grasp programming knowledge and get familiar with programming technology. 
Compared to other programming videos, the live-coding screencast is preferred because it makes programming easy to understand and demonstrates good programming habits~\cite{raj2018role}.
In practice, developers treat the source code in live-coding screencasts as important source of information and reference\cite{macleod2015code},
 but developers have difficulty in interacting with the code as they need to transcribe the code first.

While recent work has proposed to automatically extract code from live-coding screencasts to ease developer interaction with code~\cite{bao2020psc2code,ott2018deep, alahmadi2020code}, 
they obtained these live-coding screencasts either by handpicking high quality playlists of programming tutorials~\cite{bao2020psc2code} or by manually searching for live-coding screencasts~\cite{ott2018deep, alahmadi2020code}. 
These manual processes are time-consuming. The vast amount and different types of video resources on the Internet makes it difficult to search for live-coding screencasts.  As it is hard to accurately guess whether a video is a  live-coding screencast from its title and preview image, developers may need to go through many videos returned by the search result one-by-one until they find live-coding screencasts. \textcolor{blue}{}

Considering the above scenario, we develop \toolname{} to automatically identify  live-coding screencasts from a set of videos. For this identification, \toolname{} takes a number of video frame samples from the video. It then uses a classifier to determine if a video frame contains an Integrated Development Environment (IDE) window. 
Based on the frames showing IDE windows, the \toolname{} identifies the video as a  live-coding screencast. The identified live-coding screencasts can then be passed to any existing video-to-code extraction tools~\cite{bao2020psc2code, alahmadi2020code} to automatically provide the source code to developers. Therefore, \toolname{} complements these video-to-code extraction tools.

As a preliminary experiment, we collect a dataset of 112 videos sampled from the top-5 videos returned by YouTube when querying using the top-100 Java library names in Maven Repository\footnote{\url{https://mvnrepository.com/}}. Specifically, 80 live-coding screencasts covering different kinds of IDEs and 32 non live-coding screencasts are included in this dataset. Using stratified random sampling, we take 80\% of the videos as training data and 20\% of them as test data. \toolname{} achieves an F1-score of 0.97
, which is better than a random baseline 
or a baseline that identifies all videos as live-coding screencasts. 
Our analysis also found that \toolname{} is capable to identify live-coding screencasts against non live-coding screencasts which frames contain IDE screenshots.

Considering that online video platforms generate a huge amount of video every day (eg., more than 500 hours of contents are uploaded to YouTube every minute\footnote{https://blog.youtube/press/}), \toolname{} can be used by a video analysis solution to more efficiently find and process live-coding screencasts from enormous amount of fast-growing online video resources. The larger the dataset size is, the more time \toolname{} can save.
Additionally, \toolname{} can potentially be used as a browser plug-in to optimize and ease the user's experience when searching live-coding screencasts by providing only videos that are live-coding screencasts in the search result.

The contributions of this paper are:    
\begin{itemize}[nosep,leftmargin=*]
    \item We are the first to develop a tool to identify whether a video is a live-coding screencast. It eases the search of live-coding screencasts and complements existing video-to-code extraction approaches~\cite{bao2020psc2code, alahmadi2020code}.
    \item On a dataset of 112 videos with 80 live-coding screencasts and 32 non live-coding screencasts, \toolname{} achieves an F1-score of 96.97\%, which is better than a random baseline and a baseline that predicts all videos as live-coding screencasts. 
    \item Compared to the baseline, \toolname{} can shave off 32\% of running time without missing any live-coding screencasts.
\end{itemize}

The remainder of this paper is structured as follows. We describe \toolname{} in Section~\ref{sec:approach}. We then describe our experimental settings and research questions in Section~\ref{sec:exp}. We present answers to the research questions in Section~\ref{sec:results}. We present related work in Section~\ref{sec:related}. Finally, we conclude the paper and mention some future work in Section~\ref{sec:conclusion}.




\section{Approach}\label{sec:approach}\label{sec:approach}
As shown in Figure~\ref{fig-approach}, \toolname{} accepts as input a set of videos that possibly contain live-coding screencasts and outputs the set of videos classified as live-coding screencasts. Firstly, we extract sets of frames from videos and mark duplicate frames using {\em Frame Sampler}. Duplicate frames are frames that have little or no difference with their preceding frames. 
After marking the duplicate frames, we feed the non-duplicated frames into {\em Video Classifier}, which classifies whether the frames contain an IDE window. We call a frame containing an IDE window as an IDE frame.
Given the set of IDE frames from a video, {\em Video Classifier} applies a classification strategy to predict whether the video is a live-coding screencast. Finally, we return the set of videos predicted as live-coding screencasts. We describe {\em Frame Sampler} in Section~\ref{sec:sampler} and {\em Video Classifier} in Section~\ref{sec:clasifier}.

\subsection{Frame Sampler}\label{sec:sampler}
%


In most cases, analyzing all video frames is not necessary since the contents of consecutive frames are likely to be very similar or identical. Therefore, to reduce processing time, we sample the frames with a rate of one frame per 30 seconds.
Then, we mark the duplicate frames. 
To do so, we utilize the normalized root-mean-square error $ \left( NRMSE \right)$ to compute the dissimilarity of two consecutive frames on the pixel level. NRMSE score ranges from 0 to 1. The score 0 means that the two frame are identical while the score 1 means that the two frames are completely different. Given two frames $f_{i}$ and $f_{j}$, NRMSE can be computed by following formula:


\begin{equation} 
 \sqrt{\frac{\sum_{n=0}^{N-1} \sum_{m=0}^{M-1} \left [ f_{i}\left ( m,n \right )-f_{j}\left ( m,n \right ) \right ]^{2}}{\sum_{n=0}^{N-1} \sum_{m=0}^{M-1}  \left [ f_{i}\left ( m,n \right ) \right ]^{2}}} 
\end{equation}

\noindent where $f_{i}\left(m,n \right)$ and $f_{j}\left(m,n \right)$ are the pixel values at row $m$ and column $n$ for frame $f_{i}$ and $f_{j}$, respectively. $M$ and $N$ are the width and the height of the frames.
We represent a frame sequence as $\left\{ f_{i} \right\}$ where $ 0\le i\le K $, and $K$ denotes the last minute of video.
For each subsequent frame $f_{j}\left(j\ge i\ge 1\right)$ starting from $f_{i}$, we calculate the dissimilarity of $f_{i}$ and $f_{j}$ by NRMSE and delete successive frame $f_{j}$ until the dissimilarity between $f_{j}$ and $f_{i}$ is above a threshold. 
We set the threshold to 0.05 rather than 0 to account for small differences that are expected in a live-coding session  (e.g, consecutive frames change due to a mouse cursor movement or different time shown in the desktop clock).
Finally, non-duplicate frames are fed to the {\em Video Classifier}.
\subsection{Video Classifier}\label{sec:clasifier}
After getting non-duplicate frames from {\em Frame Sampler}, we apply both frame-level classifier and video-level classification strategy to classify videos into two categories: live-coding screencasts and non live-coding screencasts.

\subsubsection{Frame-Level Classifier}

We apply a deep learning model to build a frame classifier that categorizes a frame as IDE frame or non-IDE frame. IDE frame is the frame that contains an IDE window while non-IDE frame is the frame that does not contain an IDE window. 
We use the ViT~(Vision Transformer) model\cite{dosovitskiy2020image} composed of transformer encoder architecture to extract image features. ViT model is pre-trained on the ImageNet-21K dataset and has shown excellent performances on image classification tasks.
We fine tune ViT model using our labelled frames. 
By using the fine-tuned model, each non-duplicate frame is classified to either an IDE or non-IDE frame. 

\subsubsection{Video-Level Classification Strategy}
Intuitively, live-coding screencasts should continuously show IDE windows over a period of time where developers are interacting with the IDE (e.g., writing and editing the code, debugging the code, etc.). Within this period of time, the IDE frames should change their content, as opposed to displaying the same content (which may indicate that the IDE window is only a screenshot). 
Thus, \toolname{} determine a video as a live-coding screencast by following a two-stage classification strategy: (1) \toolname{} detect if there is at least one contiguous sequence of sampled frames with a minimum size of $s$ that are all non-duplicate frames and classified as IDE frames by ViT models. The size $s$ signifies the amount of changes we should observe; (2) \toolname{} detect whether the proportion of IDE frames to all non-duplicate frames exceeds a threshold $t$. A sequence of non-duplicate frames that are all classified as IDE frames signifies that the content of those frames is about IDE and changes over time. If the above two conditions are fulfilled, \toolname{} predicts the video as a live-coding screencast. 

Given sampled frames $v$ that are non-duplicate and identified as IDE frames by {\em Frame-Level Classifier},
\toolname{} searches for the shortest contiguous frame sequence
$v_{1},...,v_{j}$ where $ s\le j\le L$, and $L$ denotes the number of sampled frames. 
By default, we set $s$ to 4. 
Also, considering that the number of IDE frames in the sampled frames is $N_{ide}$ and the number of non-duplicate frames in the sampled frames is $N_{info}$, \toolname{} will compare the value of $N_{ide}/N_{info}$ with threshold $t$. 
We set $t$ to 0.5 since we want at least half of a video to contain IDE frames for it to be classified as a live-coding screencast.
For each video, if $N_{ide}/N_{info}\geq t$ and there exists an consecutive frame sequence,
$v_{1},...,v_{j}$,
in which each frame is non-duplicate and identified as IDE frame, \toolname{} predict the corresponding video as a live-coding screencast.

\section{Preliminary Experiment}\label{sec:exp}
\subsection{Dataset}


We have 112 videos in our dataset, including 80 live-coding screencasts and 32 non live-coding screencasts.
We obtain 50 live-coding screencasts from {\em psc2code}~\cite{bao2020psc2code} dataset that contain the videos with Eclipse IDE and the other 30 live-coding screencasts with other IDEs by searching on YouTube. We also obtain the 32 non live-coding screencasts from YouTube. 


We download videos from YouTube that cover live-coding screencasts of other Java IDEs and non live-coding screencasts. To find these videos, we consider a scenario where developers search for live-coding screencasts to learn how to use libraries. Specifically, we consider the top-100 libraries in Maven Repository.
For each library, we search YouTube\footnote{\url{https://www.youtube.com/}} with a query "java $\langle$library name$\rangle$". Considering a real-life scenario where a user is unlikely to go through all the search results from YouTube, we pick only the top-5 videos from each search result. From the collection of these videos, we randomly pick 100 candidates for our dataset
and download them using Pytube\footnote{\url{https://pytube.io/en/latest/}}. We manually check whether they are live-coding screencasts, non live-coding screencasts, or videos unrelated to programming (e.g., some video blogs). We find 30 live-coding screencasts, 32 non live-coding screencasts, and 38 videos that are not related to programming. The live-coding screencasts cover mainstream IDEs in the JAVA community (15 covers IntelliJ IDEA{\footnote{https://www.jetbrains.com/idea/}}, 10 covers Visual Studio Code{\footnote{https://code.visualstudio.com/}}, and 5 covers Apache NetBeans{\footnote{https://netbeans.apache.org/}}).



\begin{table}[t]
\caption{Statistics of the Videos}
\label{tab2}
\centering
\begin{tabular}{lrrr}
\cline{1-4}
      & \textbf{\# Screencast} & \textbf{\# Non-screencast} & \textbf{Total} \\ \cline{1-4}
\textbf{Train} & 64    & 25      & 89    \\
\textbf{Test}  & 16    & 7       & 23    \\ \cline{1-4}
\textbf{Total} & 80    & 32      & 112   \\ \cline{1-4}
\end{tabular}
\end{table}

To construct our labeled frame dataset, we extract one frame per second from our video dataset. 
We use the NRMSE algorithm described in Section~\ref{sec:sampler} to delete identical or highly similar frames.
Since the videos vary in length from several minutes to several hours, frames from the long videos would take a large proportion of training data if we do not limit the maximum number of frames extracted per video. 
Hence, to avoid a possible bias, we set the maximum number of extracted frames from one video to 600. If one video has more than 600 extracted frames, we randomly select the 600 frames.

Next, we label these frames to use for training our frame classifier. Frames coming from non live-coding screencasts are labeled as non-IDE frame since the non live-coding screencasts do not contain IDE. Frames of live-coding screencasts originating from {\em psc2code} are labeled with IDE-related categories: frames containing IDE window with a clear code area and frames containing an IDE window without clear code area. We relabel these frames as IDE frames as they both contain IDE window. The other frames from {\em psc2code} dataset are labeled as non-IDE frames.
Next, we manually label the frames of live-coding screencasts that are not originating from {\em psc2code} dataset. These frames are labeled by the first author. Note that labelling IDE/non-IDE frames is an objective task that can be done by a human (with no visual impairment) with perfect (or at least almost perfect) accuracy. It is clear to the labeller whether a frame contains an IDE or not. Therefore, labelling by one person is sufficient. The details of the labeled frames are shown in Table~\ref{tab1}.

\begin{table}[!htbp]
\caption{Statistics of the Labeled Frames}
\label{tab1}
\centering
\begin{tabular}{lrrr}
\cline{1-4}
& \textbf{\# IDE} & \textbf{\# Non-IDE} & \textbf{Total} \\ \cline{1-4}
\textbf{Train} & 11,741 & 11,945   & 23,686 \\
\textbf{Test}  & 2,938  & 1,511    & 4,449  \\\cline{1-4}
\textbf{Total} & 14,679 & 13,456   & 28,135 \\\cline{1-4}
\end{tabular}
\end{table}

\subsection{Experimental Settings}
We apply stratified random sampling, drawing 20\% videos from live-coding screencasts and 20\% from non live-coding screencasts, respectively. These videos are considered the testing data and remaining live-coding screencasts and non live-coding screencasts in our dataset are the training data.
The detailed split of our video dataset is shown in Table~\ref{tab2}. We use FFMPEG\footnote{\url{https://www.ffmpeg.org/}} to extract frames from the videos.
Every frame is converted into a frame of size (300$ \times $300) pixels.
We use PyTorch
to implement our deep learning model. We set the maximum number of training iterations as 100. We use Adam~\cite{DBLP:journals/corr/KingmaB14} to optimize the model. We run our experiment on a machine with an Intel Core i7 CPU, 64GB memory, and two NVidia RTX 3070 GPU with 8GB RAM.

\subsection{Research Questions}\label{sec:rqs}

\vspace{0.2cm}\noindent\textit{RQ1. Can \toolname{} effectively identify if a video is a live-coding screencast?}

\noindent We run \toolname{} and compare it with two baselines. The first baseline  randomly classifies a video into live-coding or non live-coding screencasts with 50\% chance, which we call as the {\em random baseline}. The second baseline classifies all videos as live-coding screencasts, which we call as the {\em all positive baseline}. We evaluate them using Precision, Recall, and F-1 score as defined below. We repeat the random baseline twenty times and report the average scores for each metric. 
\begin{equation} \label{equation1}
Precision = \frac{TP}{TP+FP}  
\end{equation}
\begin{equation} \label{equation2}
Recall = \frac{TP}{TP+FN}
\end{equation}
\begin{equation} \label{equation3}
F1\textit{-}score = \frac{2\times Precision\times Recall}{Precision+Recall}
\end{equation}

\vspace{0.2cm}\noindent 
TP occurs when a live-coding screencast is correctly classified as a live-coding screencast. FP occurs when a non live-coding screencast is incorrectly classified as a live-coding screencast. FN occurs when a live-coding screencast is incorrectly classified as a non live-coding screencast.

\vspace{0.2cm}\noindent\textit{RQ2: What are the cases where \toolname{} fail to correctly identify live-coding screencasts or non live-coding screencasts?}

We manually look at the misclassified videos. We extract the sampled frames from these videos to assist our analysis. We will consider two aspects: the content of informative frames in the misclassified video; and the similarity of the misclassified video with both live-coding and non live-coding screencasts.

\section{Results}\label{sec:results}

\subsection{Effectiveness in Video Classification}
Table~\ref{tab4} shows the performance of \toolname{} on identifying live-coding screencasts. 
Compared with the random baseline, \toolname{} improves recall by 92.31\%\footnote{improvement computed by: $(\frac{recall~of~\toolname{}}{recall~of~random~baseline}-1)\times 100\%$}, precision by 42.42\%, and F1 score by 67.24\%. Compared with the all positive baseline, \toolname{} has the same recall and improves precision by 28.77\%, and F1 score by 11.90\%. It indicates that \toolname{} effectively identify the live-coding screencasts. More specifically, the recall of \toolname{} is 1, which illustrates the reliability of \toolname{} in capturing live-coding screencasts.

\begin{table}[t]
\caption{Effectiveness of \toolname{}}
\label{tab4}
\centering
\begin{tabular}{@{}lrrr@{}}
\hline
             & \textbf{Recall} & \textbf{Precision} & \textbf{F1 Score} \\ \hline
Random Baseline     & 0.52 & 0.66 &0.58 \\
All Positive Baseline     & 1 & 0.73 & 0.84 \\
\toolname{}    & 1   & 0.94      & 0.97 \\
\cline{1-4}
\end{tabular}
\end{table}
\subsection{Analysis of Misclassified Videos}
\toolname{} incorrectly identifies a non live-coding screencast that contains only IDE screenshots\footnote{\url{https://www.youtube.com/watch?v=angDXZBp1zc}} as a live-coding screencast. This video is the second video returned by YouTube when searching using the keywords ``java'' and ``GPars''. It introduces various use cases of GPars library\footnote{\url{http://www.gpars.org/}} in Java. 
The developer used screenshots rather than screen recordings to demonstrate the coding process. 


Considering that the other non live-coding screencasts containing IDE screenshots in our test dataset are correctly identified, we analyse how this video differs from the others. We extracted the sampled frames from this video and found that this video shows not only the code screenshots but also a live shot of the presenter in the bottom left corner of the screen.
The IDE screenshots in these frames lead to a misclassification by our {\em frame-level classifier}. The continuous movement of the presenter's live shot causes these frames not to be considered duplicate frames despite the fact that they show the same IDE screenshot.
It appears that our {\em video-level classification strategy} is unable to handle 
a large moving objects in the frames (e.g., live shots of the user).
In the future, rather than measuring the pixel-level similarity between consecutive frames, we can potentially use OCR (optical character recognition) technologies to extract the code snippet from frames and calculate the similarity between the code snippets to avoid this kind of misclassification. 
 




\section{Related Work}\label{sec:related}
Previous work discussed the role of social media in software development~\cite{storey2014r}, which mentions the increasing role of video podcasts in software development, especially the 
practical application of video podcasts in teaching programming. 
Following~\cite{storey2014r}, researchers began to focus on how developers can share and document knowledge by using live-coding screencasts~\cite{macleod2015code}. 
By analyzing programming videos and interviewing developers, the researchers demonstrate how well live-coding screencasts can transfer knowledge between developers and build reputations for developers who created the videos.

With the rapid development of live-coding screencasts, several kinds of video-to-code tools have been proposed~\cite{yadid2016extracting,bao2020psc2code,ponzanelli2016too,khandwala2018codemotion,bao2015reverse}. Video-to-code tools leverage shape detection algorithms to find code areas and optical character recognition (OCR) technology to transcribe the code in an image to text.
ACE~\cite{yadid2016extracting} proposes a video-to-code tool which could correct OCR errors by applying statistical language models. Besides, CodeTube~\cite{ponzanelli2016too} utilizes OCR to extract code from frames of live-coding screencasts and computer vision techniques, including image segmentation and shape detection, to recognize the code area in the frame. To provide better experience, Codemotion~\cite{khandwala2018codemotion} creates an additional UI window to show the extracted code. 
After that, {\em psc2code}~\cite{bao2020psc2code} points out and eliminates noise issues that was not handled by CodeTube~\cite{ponzanelli2016too} and achieves the state-of-the-art performance.

In addition to extracting code from live-coding screencasts, recent works are also looking at other ways videos can help developers~\cite{bernal2020translating,cooper2021takes,alahmadi2020ui}. 
V2S~\cite{bernal2020translating} is a lightweight tool to obtain and capture useful video information, including bugs or feature requests from screen recordings of mobile applications. Moreover, it can translate video recordings into replayable scenarios. 
TANGO~\cite{cooper2021takes} detects duplicate video-based bug reports. TANGO
leverages both visual and textual information to detect videos that are reporting the same bug.
To improve the learning efficiency of mobile app development, UIScreens~\cite{alahmadi2020ui} could localize and extract the most representative UI screens in a mobile programming screencast, thus developers would quickly comprehend what an app displayed in a video is about.
To the best of our knowledge, while there have been some work on how videos can benefit the developer community, none of them automatically identifies live-coding screencasts. 

\balance{}
\section{Conclusion and Future Work}\label{sec:conclusion}

We propose \toolname{} to identify live-coding screencasts from online videos. From input videos, we sample a set of frames for every video and delete similar frames. Then, we feed the sets of frames into our classifier and determine whether the videos are live-coding screencasts. 
The evaluation shows that \toolname{} can identify live-coding screencast with an F1 score of 0.97. 
In the future, we plan to evaluate \toolname{} using more videos. We also plan to improve our classifications strategy by adding more rules such as learning the minimum number of IDE frames for live-coding screencasts from training videos. Moreover, we plan to utilize the information extracted from videos by \toolname{} to help automating software engineering tasks, e.g., library recommendation \cite{thung2013automated}, automated debugging and repair \cite{wang2014version,le2016history}, etc.

\vspace{0.2cm}\noindent {\bf Replication Package.} The source code for \toolname{} is available at \url{https://github.com/soarsmu/PSFinder}.


\section*{ACKNOWLEDGMENT}
This research/project is supported by the Ministry of Education, Singapore, under its Academic Research Fund Tier 2 (Award No.: MOE2019-T2-1-193). Any opinions, findings and conclusions or recommendations expressed in this material are those of the author(s) and do not reflect the views of the Ministry of Education, Singapore.

\bibliography{references}

\bibliographystyle{IEEEtran}
\end{document}